\documentclass[proceedings]{stacs}
\stacsheading{2010}{573-584}{Nancy, France}
\firstpageno{573}

\usepackage{verbatim}
\usepackage{amssymb,amsmath,amsthm}
\usepackage{algorithm, algorithmic}
\usepackage{graphicx}
\usepackage[usenames]{color}

\numberwithin{equation}{section}

\newtheorem{notation}[thm]{Notation}
\newcommand{\OPT}[1]{$\text{OPT}_{#1}^\prime$}
\newcommand{\nOPT}[1]{$\widehat{\text{OPT}_{#1}}$}

\newcommand{\cost}[1]{\text{cost}(#1)}	
\newcommand{\profit}[1]{\text{profit}(#1)}	
\newcommand{\gain}[1]{\text{gain}(#1)}	

\newcommand{\MinDisAgree}{\textsc{MinDisAgree}}
\newcommand{\MaxAgree}{\textsc{MaxAgree}}
\newcommand{\Greedy}{\textsc{Greedy}}
\newcommand{\Dense}{\textsc{Dense}}

\begin{document}
\sloppy
\title{Online Correlation Clustering}

\author[lab1]{C. Mathieu}{Claire Mathieu}
\address[lab1]{Department of Computer Science, Brown University, 115 Waterman
	Street, Providence, RI 02912}

\author[lab1,lab2]{O. Sankur}{Ocan Sankur}
\address[lab2]{Ecole Normale Sup\'erieure, 45, rue d'Ulm, 75005 Paris, France}

\author[lab1]{W. Schudy}{Warren Schudy}
\thanks{Part of this work was funded by NSF grant CCF 0728816.}


\begin{abstract}
We study the online clustering problem where data items arrive in an online fashion. The algorithm maintains a clustering of data items into similarity classes. Upon arrival of v, the relation between v and previously arrived items is revealed, so that for each u we are told whether v is similar to u. The algorithm can create a new cluster for v and merge existing clusters.

When the objective is to minimize disagreements between the clustering and the input, we prove that a natural greedy algorithm is O(n)-competitive, and this is optimal.

When the objective is to maximize agreements between the clustering and the input, we prove that the greedy algorithm is .5-competitive; that no online algorithm can be better than .834-competitive; we prove that it is possible to get better than 1/2, by exhibiting a randomized algorithm with competitive ratio .5+c for a small positive fixed constant c.
\end{abstract}
\maketitle

\section{Introduction}
	We study online correlation clustering.
	In correlation clustering \cite{bansal04, SRT02}, the input is a complete graph whose edges are  labeled either {\em positive},
	meaning similar, or {\em negative}, meaning dissimilar. The goal is to produce a clustering that
	agrees as much as possible with the edge labels. More precisely, the output is a
	clustering that maximizes the number of agreements, \textit{i.e.}, the sum of positive
	edges within clusters and the negative edges between clusters.
	Equivalently, this clustering minimizes the disagreements.
	This has applications in information retrieval, e.g. \cite{CR02,finkel08}.

	In the online setting, vertices arrive one at a time and the total number of vertices is unknown to the algorithm {\em a priori}. Upon the arrival of a
	vertex, the labels of the edges that connect this new vertex to the
	previously discovered vertices are revealed. The algorithm updates the clustering while preserving the clusters already identified 
(it is not permitted to split any pre-existing cluster).
	Motivated by information retrieval applications, this online model
	was proposed by Charikar, Chekuri, Feder and Motwani \cite{CCFM04}  (for another clustering problem).
	As in \cite{CCFM04}, our algorithms maintain {\em Hierarchical Agglomerative
	Clusterings} at all times; this is well suited for the applications of interest.

	The problem of correlation clustering was introduced by Ben-Dor et al.
	\cite{BSY99} to cluster gene expression patterns. 	
	Unfortunately, it was shown that even the offline version of correlation clustering is NP-hard \cite{SRT02,bansal04}. The following are the two approximation 
	problems that have been studied \cite{bansal04, charikar05, acn}:
	Given a complete graph whose edges are labeled positive or negative, find a
	clustering that minimizes the number of disagreements, or maximizes the number of agreements.
	We will call these problems \MinDisAgree{} and \MaxAgree{} respectively.
 Bansal et al.\ \cite{bansal04}  studied approximation algorithms both for
		minimization and maximization problems, giving a constant factor algorithm for \MinDisAgree, 
		and a {\em Polynomial Time Approximation Scheme (PTAS)} for \MaxAgree. 
		Charikar et al. \cite{charikar05} proved that \MinDisAgree{} is APX-hard and gave a 
		factor $4$ approximation. Ailon et al. \cite{acn} presented
		a randomized factor $2.5$ approximation for \MinDisAgree, which is currently the best known factor.
               The problem has attracted significant attention, with further work on several variants \cite{demaine06,charikar03,GG06,karpinski09,BSY99,joachims05, mathieu10}.

		In this paper, we study online algorithms for \MinDisAgree{} and \MaxAgree. 	
		We prove that \MinDisAgree ~ is essentially hopeless in the online setting:  the natural greedy algorithm is  $O(n)$-competitive,
		and this is optimal up to a constant factor, even with randomization (Theorem \ref{thm:greedy_bound}).
		The situation is better for \MaxAgree: we prove that the greedy algorithm is a $.5$-competitive (Theorem \ref{thm:greedy_max_bound}), but that no algorithm can be better than $0.803$ competitive ($0.834$ for randomized algorithms, see Theorem \ref{thm:max_bound}). What is the optimal competitive ratio? We prove that it is better than $.5$ by  exhibiting an algorithm with competitive ratio $0.5+\epsilon_0$ where
		$\epsilon_0$ is a small absolute constant (Theorem \ref{corollary:overallalgorithm}). Thus Greedy is not always the best choice!


\medskip

	More formally, let $v_1, \ldots, v_n$  denote the sequence of vertices of the input graph, where $n$ is not 
	known in advance. Between any two vertices, $v_i$ and $v_j$ for $i \neq j$, 
	there is an edge labeled positive or negative. 	
	In \MinDisAgree{} (resp. \MaxAgree), the goal is to find a clustering $\mathcal{C}$, \textit{i.e.} a partition of the nodes,
	that minimizes the number of disagreements $\cost{\mathcal{C}}$: the number of negative edges within clusters plus the 
	number of positive edges between clusters (resp. maximizes the number of
	agreements $\profit{\mathcal{C}}$: the number of positive edges within clusters plus the number of
	negative edges between clusters). Although these problems are equivalent
	in terms of optimality, they differ from the point of view of approximation.	
	Let OPT denote the optimum solution of \MinDisAgree{} and of
	\MaxAgree. 

	In the online setting, upon the arrival of a new vertex, the algorithm updates the current clustering: it may either create a
	new singleton cluster or add the new vertex to a pre-existing cluster, and may decide to merge some pre-existing clusters. It is not allowed to split pre-existing clusters.

	A $c$-competitive algorithm for \MinDisAgree{} 
	outputs, on any input $\sigma$, a clustering $\mathcal{C}(\sigma )$ such that $\cost{\mathcal{C} (\sigma )} \leq c \cdot \cost{\text{OPT}(\sigma )}$. For \MaxAgree{}, we must have
	$\profit{\mathcal{C} (\sigma )} \geq c \cdot \profit{\text{OPT}(\sigma )}$. (When the algorithm is randomized, this must hold in expectation).

\section{Maximizing Agreements Online}
\label{section:maximization}
	\subsection{Competitiveness of \Greedy{}}
	For subsets of vertices $S$ and $T$ we define $\Gamma(S, T)$ as the
	set of edges between $S$ and $ T$. We write
	$\Gamma^+(S, T)$ (resp. $\Gamma^-(S, T)$) for the set of positive
	(resp. negative) edges of $\Gamma(S,T)$.
	We define the {\em gain} of merging $S$ with $T$ 
	as the change in the profit when clusters $S$ and $T$ are merged:
	\[
		\gain{S, T} = {|\Gamma^+(S, T)| - |\Gamma^-(S, T)|}= 2|\Gamma^+(S, T)| -|S||T|.
	\]
	We present the following greedy algorithm for online correlation clustering. 	

	\begin{algorithm}[H]
		\caption{Algorithm \Greedy}
		\label{algorithm:greedy}
		\begin{algorithmic}[1]
		\FORALL{vertex $v$}
			\STATE Put $v$ in a new cluster \label{greedy:new_cluster}
			consisting of $\{v\}$.
			\WHILE{there are two clusters $C$, $C'$ such that $\gain{C, C'} > 0$}
				\STATE Merge $C$ and $C'$ \label{greedy:merge}
			\ENDWHILE
		\ENDFOR
		\end{algorithmic}
	\end{algorithm}

		\begin{thm}\label{thm:greedy_max_bound}Let OPT denote the offline optimum.
\begin{itemize}
 \item 
			For every instance, $\profit{\Greedy{}}\geq {0.5 ~\profit{\text{OPT}}}$.
\item			There are instances with $\profit{\Greedy{}}\leq {(0.5 + o(1))\profit{\text{OPT}}}$.
\end{itemize}		\end{thm}

\subsection{Bounding the optimal competitive ratio}
	\begin{thm}
		\label{thm:max_bound}
		The competitive ratio of any randomized online algorithm for \MaxAgree{} 
		is at most $0.834$.
		The competitive ratio of any deterministic online algorithm for \MaxAgree{} 
		is at most $0.803$.
	\end{thm}
	The proof uses Yao's Min-Max Theorem \cite{borodin98} (maximization version). 

	\begin{thm}[Yao's Min-Max Theorem]
		\label{thm:yao}
		Fix a distribution $D$ over a set of inputs $(I_\sigma)_\sigma$.  The competitive ratio of any randomized online algorithm is at most
		\[\max \{ \frac{E_{I}[\profit{\mathcal{A}(I)}]}{E_{I}[\profit{\text{OPT}(I)}]} :
 \mathcal{A} \hbox{ deterministic online algorithm} \} ,\]
		where the expectations are  over a random input $I$ drawn from distribution $D$.
	\end{thm}

	To prove Theorem 2.2, we first define two generic inputs that we will
	use to apply Theorem 2.3. The first input is a graph $G_1$ with $2m$
	vertices and all positive edges between them 
	The second input is a graph with $6m$ vertices defined as
	follows. The first $2m$ vertices have all positive edges between them,
	the next $2m$ vertices have all positive edges between them, and the
	last $2m$ vertices also have all positive edges between them. In each of
	these three sets $G_1,G_2,G_3$ of $2m$ vertices, half are labelled
	``left side" vertices and the other half are labelled "right side"
	vertices. All edges between left vertices are positive, but edges
	between a vertex $u$ on the left side of $G_i$ and a vertex $v$ on the
	right side of $G_j$, $j\neq i$, are all negative. 

	The online algorithm cannot distinguish between the two inputs until
	time $2m+1$, so it must hedge against two very different possible
	optimal structures. 

	\subsection{Beating \Greedy{} }

\subsubsection{Designing the algorithm}

	Our algorithm is based on the observation that Algorithm \Greedy{} always satisfies at least half of the edges.
	Thus, if \profit{OPT} is less than $(1-\alpha/2) |E|$ for some constant $\alpha$,  then the profit of \Greedy{} is better than half of optimal. 
	We design an algorithm called \Dense{}, parameterized by constants
	$\alpha$ and $\tau$, such that if \profit{OPT} is greater than
	$(1-\alpha/2)|E|$, then the approximation factor is at least $0.5+\eta$ for some positive constant $\eta$.
	We use both algorithms \Greedy{} and \Dense{} to define Algorithm \ref{overallalgorithm}.
	
	\begin{thm}
		\label{theorem:dense}
		Let $\alpha \in (0,1)$, $\tau >1$ and $\eta \in (0,\frac{1}{2})$ be such that
		\begin{equation}
			\label{eqn:constants_condition}
			\eta \leq 1.5 - \tau^2 - ((2\sqrt{3}+9/2)\alpha^{1/4} + \frac{\alpha^{1/4} }{1-\alpha^{1/4}}+ \alpha/2)2\frac{2\tau -1}{(\tau-1)}.
		\end{equation}
		Then, for every instance such that $\text{OPT} \geq (1-\alpha/2)E$, Algorithm $\Dense_{\alpha,\tau}$ has
		profit at least $(1/2 + \eta)\text{OPT}$.
	\end{thm}

	Using Theorem \ref{theorem:dense} we can bound the competitive ratio of Algorithm \ref{overallalgorithm}.

	\begin{corollary}
		\label{corollary:overall_comp_ratio}
		Let $\alpha, \tau$ and $\eta$ be as above, and let $p ={\alpha}/({2 + 2\eta(2-\alpha)})$. Then Algorithm \ref{overallalgorithm} has competitive ratio at least 
		$\frac{1}{2} + \frac{\alpha \eta/2}{1+2\eta(1-\alpha/2)}$.
	\end{corollary}

	\begin{corollary}
		\label{corollary:overallalgorithm}
		For $\alpha = 10^{-12}$, $\tau = 1.0946$, $\eta =0.0555$ and $p=4,5 \cdot 10^{-13}$,
		Algorithm \ref{overallalgorithm} is	$\frac{1}{2} + 2\cdot 10^{-14}$-competitive.
	\end{corollary}
	
	\begin{algorithm}
		\caption{A $\frac{1}{2} + \epsilon_0$-competitive algorithm}
		\label{overallalgorithm}
		\begin{algorithmic}
			\STATE Given $p$, $\alpha$, $\tau$,
			\STATE With probability $1-p$, run \Greedy{},
			\STATE With probability $p$, run $\text{\Dense{}}_{\alpha, \tau}$.
		\end{algorithmic}
	\end{algorithm}

	\begin{algorithm}
		\caption{Algorithm $\Dense{}_{\alpha,\tau}$} 
		\label{algorithm:dense}
		\begin{algorithmic}[1]
			\STATE Let $\mathcal{C} =\widehat{\text{OPT}}_{1}$ and for every cluster $D \in \mathcal{C}$, let $\text{repr}_1(D) := D
				\in \widehat{\text{OPT}}_{1}$ .
				\label{dense:OPT1}
			\FORALL{a vertex $v$ at time $t$}
				\STATE Put $v$ in a new cluster $\{v\}$.
				\IF{$t = t_i$ for some $i$}
					\FOR{every cluster $D$ in $\widehat{\text{OPT}_{i}}$}
						\label{dense:outer_loop}
						\STATE Define  a cluster $D''$ obtained by merging the restriction of $D$ to $\{t_{i-1}, \ldots, t_i\}$ with every cluster $C \in \mathcal{C}$ in $\{1, \ldots, t_{i-1}\}$ such that $\text{repr}_{i-1}(C)$ is defined and is half-contained in $D$.
						\STATE If $D''$ is not empty, set $\text{repr}_i(D'') := D \in \widehat{\text{OPT}_{i}}$.
							\label{dense:repr}
					\ENDFOR
				\ENDIF
			\ENDFOR
		\end{algorithmic}
	\end{algorithm}

	How do we define algorithm \Dense{}? Using the PTAS of \cite{bansal04}, one can compute offline a factor
	$(1-\alpha/2)$ approximative solution \OPT{}  of any instance of \MaxAgree{}
	in polynomial time.  We will design algorithm \Dense{} so that it guarantees an approximation factor of $0.5
	+ \eta$ whenever $\profit{\text{\OPT{}}} \geq (1-\alpha)|E|$. Since $\profit{\text{OPT}} \geq (1-\alpha/2)|E| $ implies that $\profit{\text{\OPT{}}} \geq (1-\alpha)|E|$, Theorem \ref{theorem:dense} will follow.

We  say that \OPT{t} is {\em large} if $\profit{\text{\OPT{t}}} \geq (1-\alpha)|E|$.
	We define a sequence $(t_i)_i$ of {\em update times} inductively as follows: By convention $t_0=0$. 
	Time $t_1$ is the earliest time $t\geq 100$  
	such that \OPT{t} is large. Assume $t_i$ is already defined, and let $j$ be such that $\tau^{j-1} \leq t_i < \tau^{j}$. 
	If \OPT{\tau^j} is large, then $t_{i+1}=\tau^j$, else $t_{i+1}$ is the
	earliest time $t\geq \tau^j$ such that \OPT{t} is large. 
	Let $t_1,t_2,\ldots ,t_K$ be the resulting sequence. We will note, with an abuse of notation, \OPT{i} instead of \OPT{t_i} for $1 \leq i \leq K$.

	We say that a cluster $A$ is {\em half-contained} in $B$ if  $|A \cap B| > |A|/2$.
	Let $\epsilon = \alpha^{1/4}$.
	For each $t_i$, we inductively define a near optimal clustering
	 of the nodes $[1,t_i]$. For the base case, let
	$\widehat{\text{OPT}}_1$ be the clustering obtained from \OPT{1} by keeping
	the $1/\epsilon^2$ largest clusters and splitting the other clusters into
	singletons. For the general case, to define \nOPT{i}  given \nOPT{i-1}, mark the
	clusters of \OPT{i} as follows. For any $D$ in \OPT{i}, mark $D$ if
either one of the $1/\epsilon^2 - 1/\epsilon$ largest clusters of
			\nOPT{i-1} is half-contained in $D$, or  $D$ is one of the $1/\epsilon$ largest clusters \OPT{i}. Then \nOPT{i} contains all the marked clusters of \OPT{i} and the rest of the vertices 
	in $[1,t_i]$ as singleton clusters.
(Note that, by definition, any \nOPT{i} contains at most $1/\epsilon^2$ non-singleton clusters; this will be useful in the analysis.)

Note that \Dense{} only depends on
parameters $\alpha$ and $\tau$ indirectly via the definition of update times
and of \nOPT{}.


	\subsubsection{Analysis: Proof of Theorem \ref{theorem:dense}}

The analysis is by induction on $i$, assuming that we start from clustering $\widehat{\text{OPT}_i}$ at time $t_i$, then apply the above algorithm from time $t_i$ to the final time $t$. If $i=1$ this is exactly our algorithm, and if $i=K$ then this is simply $\widehat{\text{OPT}_K}$; in general it is a mixture of the two constructions.

		More formally, define a forest ${\mathcal{F}}$ (at time $t$) with one node for each $t_i\leq t$ and cluster of $\widehat{\text{OPT}_i}$. The node associated to a cluster $A$ of  $ \widehat{\text{OPT}_{i-1}}$ is a child of the node associated to a cluster $B$ of $\widehat{\text{OPT}_i}$ if and only if  $A$ is half-contained in $B$. With a slight abuse of notation, we define the following clustering $\mathcal{F}$ associated to the forest. There is one cluster $T$ for each tree 
		of the forest: for each node $A$ of the tree, if $i$ is such
		that $A\in\widehat{\text{OPT}_i}$, then cluster $T$ contains $A\cap (t_{i-1},t_i]$. This defines $T$. 

\begin{figure}
			\centering
			\includegraphics{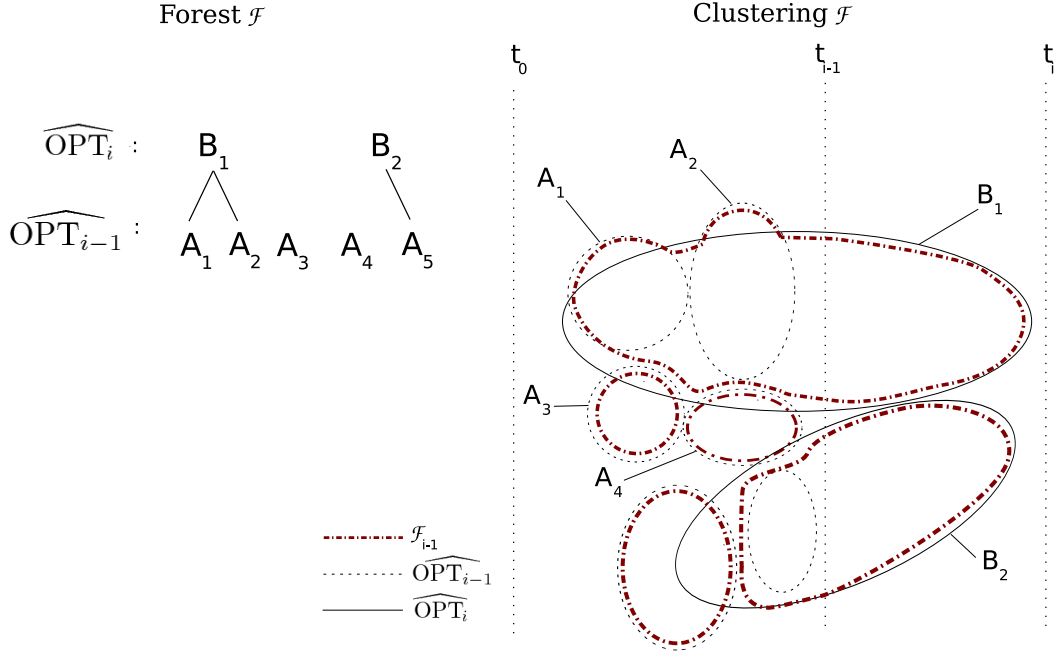}
			\caption{An example of a forest $\mathcal{F}$ given in left, and the corresponding clustering given in right. Here, we have
					$\text{\nOPT{i}}=\{B_1, B_2\}$ and $\text{\nOPT{i-1}}=\{A_1, \ldots, A_5\}$.
				}
			\label{figure:F_example}
		\end{figure}

		One interpretation of \Dense{} is that at all times $t$, there is an
		associated forest and clustering $\mathcal{F}$; and our algorithm \Dense{} simply
		maintains it.
		See Figure \ref{figure:F_example} for an example.
				

		\begin{lemma}
			\label{lemma:dense=F}
                        Algorithm~\ref{algorithm:dense} is an online algorithm that outputs clustering $\mathcal{F}$ at time $t$.
		\end{lemma}

Let ${\mathcal{F}}_i$ be the forest obtained from ${\mathcal F}$ by erasing every node associated to  clusters of $\widehat{\text{OPT}_j}$ for every $j<i$. 
		With a slight abuse of notation, we define the following clustering ${\mathcal F}_i$ associated to that forest: 
		there is one cluster $C$ for each tree of the forest defined as follows. 
		For each node $A$ of the tree, let $k\geq i$ be such that $A\in\widehat{\text{OPT}_k}$: 
		then $C$ contains $A\cap (t_{k-1},t_k]$ if $k>i$, and $C$ contains $A$ if $k=i$.
		This defines a sequence of clusterings such that ${\mathcal F}_1={\mathcal F}$ is the output of the algorithm, and ${\mathcal F}_K=\widehat{\text{OPT}_K}$.

		\begin{lemma}[Main lemma]
			\label{lemma:delta_cost}
			For any $2 \leq i \leq K$,
			\begin{equation*}
				\cost{ {\mathcal F}_{i-1}}-\cost{{\mathcal F}_{i}}
				\leq \left((4 +2\sqrt{3})\epsilon + \frac{\epsilon}{1-\epsilon} \right)t_it_K.
			\end{equation*}
		\end{lemma}
		
		We defer the proof of Lemma \ref{lemma:delta_cost} to next section.
		Assuming Lemma \ref{lemma:delta_cost}, we upper-bound the cost of clustering $\mathcal{F}$.

		\begin{lemma}[Lemma 14, \cite{bansal04}]
			\label{lemma:cost_of_splitting}
			For any $0<c<1$ and clustering $\mathcal{C}$, let $\mathcal{C}'$ be the clustering obtained from 
			$\mathcal{C}$ by splitting all clusters of $\mathcal{C}$ of size less than $cn$, where $n$ is the
			number of vertices. Then $\cost{\mathcal{C}'} \leq \cost{\mathcal{C}} + c n^2/2$.
		\end{lemma}

		\begin{lemma}
			\label{lemma:cost_F}
				$\cost{\mathcal{F}} \leq ((2\sqrt{3}+9/2)\epsilon + \frac{\epsilon}{1-\epsilon}+\epsilon^4/2)
\frac{2\tau-1}{\tau-1}t_K^2$.
		\end{lemma}
		\begin{proof}
			We write:
 $\cost{\mathcal{F}}= \cost{\widehat{\text{OPT}_K} } + 
				\sum_{i=2}^{K} (\cost{\mathcal{F}_{i-1}} - \cost{ {\mathcal{F}}_{i}})
$.%
			By definition, \nOPT{K} contains the $1/\epsilon$ largest clusters of \OPT{K}. Then
			the remaining clusters of \OPT{K} are of size at most $\epsilon t_K$.
                        By Lemma \ref{lemma:cost_of_splitting}, the cost of \nOPT{K} is at most $\cost{\text{\OPT{K}}} + \epsilon t_K^2/2 \leq {(\alpha + \epsilon)t_K^2/2}$.
			Applying Lemma \ref{lemma:delta_cost}, and summing over $2 \leq i \leq K$, we get
			\[
				\cost{\mathcal{F}} \leq {(\alpha + \epsilon)t_K^2/2} + \left((4+2\sqrt{3})\epsilon + \frac{\epsilon}{1 - \epsilon}\right) \sum_i{t_i}t_K.
			\]
			By definition of the update times $(t_i)_i$, for any $j >0$ there
			exists at most one $t_i$ such that $\tau^j \leq t_i < \tau^{j+1}$.
			Let $L$ be such that $\tau^L \leq t_K < \tau^{L+1}$. Then
$$\sum_{1\leq i \leq K}{t_i} \leq \sum_{1 \leq i \leq K-1}{t_i} + t_K
					\leq \sum_{1 \leq j \leq L}{\tau^j} + t_K
					\leq \frac{\tau^{L+1}}{\tau -1} + t_K
					\leq \frac{2\tau - 1}{\tau -1}t_K.	$$
			Hence the desired bound on $\cost{\mathcal{F}}$.
		\end{proof}


		\begin{proof}[Proof of Theorem \ref{theorem:dense}]
			Fix an input graph of size $n$, such that $\profit{\text{OPT}} \geq (1-\alpha/2){n \choose 2}$.
			By Lemma \ref{lemma:cost_F}, at time $t_K$, Algorithm \ref{algorithm:dense} has clustering $\mathcal{F}$
			with $\cost{\mathcal{F}} \leq O(\epsilon) \frac{2\tau-1}{\tau-1} t_K^2.$

			By definition of the update times, $n < \tau t_K$. 
			To guarantee a competitive ratio of $0.5 + \eta$, for some $\eta$, the cost must not exceed
			$(0.5-\eta){n \choose 2}$ at time $n$, when all vertices $t_K+1, \ldots, n$ are added as singleton clusters.
			The number of new edges added to the graph between times $t_K$ and $n$ is ${n - t_K \choose 2} + t_K (n - t_K)$.
			We must have
			\begin{equation}
				\label{equation:cost_constraint}
				\frac{2\tau-1}{\tau-1}O(\epsilon)t_K^2
				+
				{n - t_K \choose 2} + t_K (n - t_K)
				\leq (0.5 - \eta){n \choose 2},
			\end{equation}
			for some $0<\eta<0.5$. Using the fact that $n-t_K \leq (\tau-1)t_K$ and $t_K \leq n-1$, 
			to satisfy (\ref{equation:cost_constraint}), it suffices to have
			\[
				\frac{2\tau-1}{\tau-1}O(\epsilon)t_K^2
				+
				t_K^2(\tau-1)^2/2 + (\tau-1) t_K^2 
				\leq (0.5 - \eta)t_K^2/2,
			\]
			which is equivalent to (\ref{eqn:constants_condition}).
			Moreover we have the following natural constraints on constants
			$\eta$, $\epsilon$ and $\tau$: $0 < \eta < 0.5$, $0 < \epsilon < 1$, and $\tau >1$.
			Then, for any set of values of constants $\eta$, $\epsilon$, $\tau$ verifying
			those constraints, Algorithm \Dense{} is $0.5+\eta$-competitive.
		\end{proof}

	\subsubsection{The core of the analysis: proof of Lemma \ref{lemma:delta_cost}}

		\begin{lemma}
			\label{lemma:Si}
Let $\mathcal{S}^i$ be the set of vertices of the non-singleton clusters 
			that are not among the $1/\epsilon^2 - 1/\epsilon$ largest clusters of \nOPT{i-1}. Then
			$|\mathcal{S}^i| \leq \frac{\epsilon}{1-\epsilon}t_{i-1}.$
		\end{lemma}
		\begin{proof}
			Let $C$ be a cluster of \nOPT{i-1}, such that $C \subseteq \mathcal{S}^i$.
			Then ${|C| \leq (1/\epsilon^2-1/\epsilon)^{-1}t_{i-1}}$. Since there are at
			most $1/\epsilon$ such clusters, the number of vertices 
			of these are at most $1/\epsilon(1/\epsilon^2 - 1/\epsilon)^{-1}t_{i-1}$.
		\end{proof}

		\begin{notation}
			For any $i \neq j$, and a cluster $B$ of \OPT{i}, we denote by
			$\gamma^{i,j}_B$ the square root of the number of edges of 
			$[1,t_{\min(i,j)}]\times [1,t_{\min(i,j)}]$, adjacent to at least one node of $B$, 
			and which are classified differently in \OPT{i} and in \OPT{j}.
		\end{notation}

		We refer to non singleton clusters as {\em large} clusters.

		\begin{lemma}
			\label{lemma:Ti}
Let $\mathcal{T}^i$ be the set of vertices of those $1/\epsilon^2 -
			1/\epsilon$ largest clusters of \nOPT{i-1} that
			are not half-contained in any cluster of \OPT{i}. Then
$|\mathcal{T}^i| \leq \sqrt{6}\sum_{\text{large } C \in \text{\nOPT{i-1}}}{\gamma^{i,i-1}_C}
$.		\end{lemma}

		Let $B$ be a cluster of \nOPT{i}. For any $j\leq i$, we define $\mathcal{C}_j(B)$ 
		as the cluster associated with the tree of $\mathcal{F}_j$ that contains $B$. 
		For any $B \in \text{\nOPT{i}}$, we call
		$\mathcal{C}_{i-1}(B)$ the {\em extension} of $\mathcal{C}_i(B)$ to $\mathcal{F}_{i-1}$.
		By definition of $\mathcal{F}_i$, the following lemma is easy. 
		\begin{lemma}
			\label{lemma:restriction_of_Ci}
			For any $B \in \text{\nOPT{i}}$, the restriction of $\mathcal{C}_{i-1}(B)$ to $(t_{i-1},t_K]$ is equal to the restriction of $\mathcal{C}_i(B)$ to $(t_{i-1},t_K]$.
		\end{lemma}

		Let $(A_j)_j$ denote the clusters of \nOPT{i-1} that are half-contained in $B$.
		We define $\delta^i(B)$ as the symmetric difference of the restriction
		of $B$ to $[1,t_{i-1}]$ and $\cup_j{A_j}$: 
		\[
			\delta^i(B) = (B \cap [1,t_{i-1}]) \Delta \cup_j{A_j}.
		\]

		\begin{lemma}
			\label{lemma:inclusion}
			For any cluster $C_i$ of $\mathcal{F}_i$, let $C_{i}'$ denote the extension of $C_i$ to $\mathcal{F}_{i-1}$. Then
			\[
				\bigcup_{C_i \in \mathcal{F}_i}{{C}_i \setminus {C}_{i}'}
				\subseteq \mathcal{S}^i \cup \mathcal{T}^i \cup
					\bigcup_{\text{large }B \in \text{\nOPT{i}}}{\delta^i(B)}
			\]
		\end{lemma}
		\begin{proof}
			By Lemma \ref{lemma:restriction_of_Ci},
			the partition of the vertices $(t_{i-1}, t_K]$ is the same for $C_i$ as for $C_{i}'$.
			So $C_i$ and $C_{i}'$ only differ in the vertices of $[1,t_{i-1})$:
			\[
				\bigcup_{C_i \in \mathcal{F}_i}{{C}_i \setminus {C}_{i}'}
				\subseteq \bigcup_{B \in \text{\nOPT{i}}}{\delta^i(B)}.
			\]

			We will show that for a singleton cluster $B$ of \nOPT{i}, $\delta^i(B)$
			is included in $\mathcal{S}^i \cup \mathcal{T}^i \bigcup_{\text{large }B \in \text{\nOPT{i}}}{\delta^i(B)}$, 
			which yields the lemma.

			Let $B = \{v\}$ be a singleton cluster of \nOPT{i} such that $\delta^i(B) \neq \{\}$.
                        A non-singleton cluster cannot be half-contained in a singleton cluster so we conclude no clusters are half-contained in $B$ and hence $\delta^i(B) = \{v\}$.
                        By definition of $\delta^i(B)$, $v \in [1,t_{i-1}]$. So there
			exists a cluster $A$ of \nOPT{i-1} that contains $v$. Clearly $A$ is not a singleton since otherwise $\delta^i(B)$ would be $\{\}$.			
			There are two cases.

 First, if $A$ is half-contained in a cluster $B' \neq B$ of
				\nOPT{i} then cluster $B'$ is necessarily large since it contains
				more than one vertex of $A$. Then we have $v \in \delta^i(B')$.

 Second, if $A$ is not half-contained in any cluster of \nOPT{i}
				then $A \subseteq \mathcal{S}^i \cup \mathcal{T}^i$. In fact,
				if $A$ is half-contained in a cluster of \OPT{i} which is split
				into singletons in \nOPT{i}, then $A$ is not one of the
				$1/\epsilon^2 -1/\epsilon$ largest clusters of \nOPT{i-1},
				and $A \subseteq \mathcal{S}^i$. If $A$ is not half-contained
				in any cluster of \OPT{i}, then $A \subseteq \mathcal{T}^i$ if
				$A$ is one of the $1/\epsilon^2 - 1/\epsilon$ largest clusters of \nOPT{i-1} and $A \subseteq
				\mathcal{S}^i$ otherwise.
		\end{proof}

		\begin{lemma}
			\label{lemma:structure_mixed_clusterings}
			For any large cluster $B$ of \nOPT{i}, $|\delta^i(B)| \leq 2\sqrt{2}\gamma^{i,i-1}_B$.
		\end{lemma}
		\begin{proof}
			Let $B'$ denote the restriction of $B$ to $[1,t_{i-1}]$.
			We first show that 
			\[
				1/2(|\cup_j{A_j} \setminus B'|)^2 \leq (\gamma^{i,i-1}_B)^2.
			\]
			Observe that $(\gamma^{i,i-1}_B)^2$ includes all edges $uv$ such
			that one of the following two cases occurs.

First, if $u\in A_j\setminus B$ and $v\in A_j\cap B$: such edges are
					internal in the clustering \OPT{i-1} but external in the
					clustering \OPT{i}. The number of edges of this type is 
					$\sum_j |A_j\setminus B|\cdot |A_j\cap B|$. 
					Since $A_j$ is half-contained in $B$, 
					this is at least $ \sum_j |A_j \setminus B|^2$.

Second, if $u\in A_j \cap B$ and $v\in A_k\cap B$ with $j\neq k$: 
					such edges are external in the clustering \OPT{i-1} but internal in 
					the clustering \OPT{i}. The number of edges of this type is 
					$\sum_{j<k} |A_j\cap B|\cdot |A_k \cap B| \ge \sum_{j<k} |A_j\setminus B|\cdot |A_k \setminus B|$.

 			Summing, it is easy to infer that $(\gamma^{i,i-1}_B)^2\geq (1/2)\left(\sum_j |A_j \setminus B|\right)^2
					=(1/2)|\cup_j A_j \setminus B'|^2
$.			Let $(A'_j)_j$ denote the clusters of \nOPT{i-1} that are not
			half-contained in $B$, but have non-empty intersections with $B$.
			We now show that 
			\[
				1/2(|B' \setminus \cup_j{A'_j}|)^2\leq (\gamma^{i,i-1}_B)^2.
			\]
			We have $B' \setminus \cup_j{A_j} = \cup_j{(A'_j \cap B)}$.
			Observe that any $A'_j$ is a large cluster of \nOPT{i-1},
			thus a cluster of \OPT{i-1}. Then $(\gamma^{i,i-1}_B)^2$ includes all 
			edges $uv$ such that one of the following two cases occurs

First, if $u\in A'_j\setminus B$ and $v\in A'_j\cap B$: such edges are
					internal in the clustering \OPT{i-1} but external in the
					clustering \OPT{i}. The number of edges of this type is 
					$\sum_j |A'_j\setminus B|\cdot |A'_j\cap B|$. 
					Since $A'_j$ is not half-contained in $B$, 
					this is at least $\sum_j |A'_j \cap B|^2$.

Second, if $u\in A'_j \cap B$ and $v\in A'_k\cap B$ with $j\neq k$: 
					such edges are external in the clustering \OPT{i-1} but internal in 
					the clustering \OPT{i}. The number of edges of this type is 
					$\sum_{j<k} |A'_j\cap B|\cdot |A'_k \cap B|$.

 			Summing, we get
			\[
				(\gamma^{i,i-1}_B)^2\geq (1/2)\left(\sum_j |A'_j \cap B|\right)^2
					=(1/2)|B' \setminus \cup_j A'_j|^2.
			\]
		\end{proof}

		\begin{lemma}
			\label{lemma:structure1}
			 For any $i\geq 1$, \nOPT{i} has at most $1/\epsilon^2$ non singleton clusters, all
			 of which are clusters of \OPT{i}
		\end{lemma}
		\begin{proof}
				By definition, \nOPT{1} has at most $1/\epsilon^2$ non singleton
				clusters.
				For any $i > 1$, a cluster of \nOPT{i-1} can only be half-contained in one cluster of
				\OPT{i}. Therefore given \nOPT{i-1}, at most $1/\epsilon^2$ clusters of
				\OPT{i} are marked. Thus \nOPT{i} has at most $1/\epsilon^2$ clusters.
		\end{proof}

		We can now prove Lemma \ref{lemma:delta_cost}.

		\begin{proof}[Proof of Lemma \ref{lemma:delta_cost}]
			By Lemma \ref{lemma:restriction_of_Ci}, clusterings $\mathcal{F}_i$ and $\mathcal{F}_{i-1}$ only
			differ in their partition of $[1,t_{i-1}]$.	Then the set of the vertices that are classified 
			differently in $\mathcal{F}_i$ and $\mathcal{F}_{i-1}$
			is $\cup_i C_i \setminus C_{i-1}$. Each of these vertices creates at most $t_K$ disagreements:
			\begin{equation}
			\begin{split} \label{eqn:delta_cost}
				\cost{ {\mathcal F}_{i-1}}-\cost{{\mathcal F}_{i}}
				&\leq \sum_{C_i \in \mathcal{F}_i}{|C_i \setminus C_{i-1}|}t_K\\
			\end{split}
			\end{equation}
			By Lemmas \ref{lemma:inclusion} and \ref{lemma:structure_mixed_clusterings},
			\begin{equation} \label{eqn:delta_cost0}
				\sum_{C_i \in \mathcal{F}_i}{|C_i \setminus C_{i-1}|}t_K
				\leq \left( 2\sqrt{2} \Bigg(\sum_{\text{large } B\in \widehat{\text{OPT}_i}} 
						\gamma^{i,i-1}_B\Bigg) + |\mathcal{S}^i| + |\mathcal{T}^i|\right) t_K.
			\end{equation}
			By Lemmas \ref{lemma:Si} and \ref{lemma:Ti},
			\begin{equation}
				\label{eqn:delta_cost1}
				|\mathcal{S}^i| \leq \frac{\epsilon}{1-\epsilon} t_{i-1} \hbox{ and } 				|\mathcal{T}^i| \leq \sqrt{6}\sum_{\text{large }B \in \text{\nOPT{i-1}}}{\gamma^{i-1,i}_B}
			\end{equation}
			The term $\sum_{\text{large } B\in \widehat{\text{OPT}_{i-1}}} \gamma^{i-1,i}_B$ can be seen as the
			$\ell_1$ norm of the vector $(\gamma^{i-1,i}_B)_{\text{large } B}$. Since \nOPT{i-1} has at most 
			$1/\epsilon^2$ large clusters by Lemma \ref{lemma:structure1}, we can use H\"older's inequality:
			\begin{equation*}
				\begin{split}
					\sum_{\text{large } B\in \widehat{\text{OPT}_{i-1}}} \gamma^{i-1,i}_B
					&= \Vert (\gamma^{i-1,i}_B)_{\text{large } B} \Vert_1 \leq 1/\epsilon \Vert (\gamma^{i-1,i}_B)_{\text{large } B} \Vert_2.
				\end{split}
			\end{equation*}
			By definition we have $\Vert
			(\gamma^{i-1,i}_B)_{\text{large B}}
			\Vert_2 \leq \sqrt{2(\cost{\text{\OPT{i-1}}} + \cost{\text{\OPT{i}}})}$. Thus
			\begin{equation}
				\label{eqn:delta_cost4}
				\sum_{\text{large } B\in \widehat{\text{OPT}_{i-1}}} \gamma^{i-1,i}_B
				\leq  1/\epsilon \sqrt{2(\alpha t_{i-1}^2/2 +
						\alpha t_i^2/2)} \leq \frac{\sqrt{2\alpha}}{\epsilon}t_i.
			\end{equation}
			Similarly, we have
			\begin{equation}
				\label{eqn:delta_cost5}
				\sum_{\text{large } B \in \text{\nOPT{i}}}{\gamma^{i,i-1}_B}
				\leq \frac{\sqrt{2\alpha}}{\epsilon}t_i.
			\end{equation}
			Combining equations (\ref{eqn:delta_cost}) through (\ref{eqn:delta_cost5}) and $\alpha=\epsilon^4$ yields
			\[
				\cost{ {\mathcal F}_{i-1}}-\cost{{\mathcal F}_{i}}
				\leq \left( (4 +2\sqrt{3})\epsilon + \frac{\epsilon}{1-\epsilon} \right) t_i t_K
			\]
		\end{proof}

\section{Minimizing Disagreements Online}
	\label{section:minimization}

	\begin{thm}
		\label{thm:greedy_min}
		Algorithm \Greedy{} is $(2n+1)$-competitive for \MinDisAgree.
	\end{thm}

	To prove Theorem \ref{thm:greedy_min}, we need to compare the cost of the optimal clustering to the cost of the clustering constructed by the algorithm. The following lemma reduces this to, roughly, analyzing the number of vertices classified differently. 

	\begin{lemma}
		\label{lemma:greedy_WW'}
		Let $\mathcal{W}$ and $\mathcal{W}'$ be two clusterings such that there is an injection
		$W'_i \in \mathcal{W}' \rightarrow W_i \in \mathcal{W}$.
		Then $\cost{\mathcal{W}'} - \cost{\mathcal{W}} \leq n \sum_i |W_i' \setminus W_i|$.
	\end{lemma}

	For subsets of vertices $S_1, \ldots, S_m$, we will write, with a slight
	abuse of notation, $\Gamma^+(S_1, \ldots, S_m)$ for the set of edges in
	$\Gamma^+(S_i, S_j)$ for any $i \neq j$: $\Gamma^+(S_1, \ldots, S_m) = \cup_{i \neq j}{\Gamma^+(S_i, S_j)}$.

	\begin{lemma}
		\label{lemma:greedy_inequality}
		Let $C$ be a cluster created by \Greedy, and $\mathcal{W} =\{W_1, \ldots, W_K\}$ denote the clusters of OPT.
Then $			|C| \leq \max_i |C \cap W_i| + 2|\Gamma^+(C \cap W_1, \ldots, C \cap W_K)|.
$.		We call $\displaystyle i_0= \arg\max_i |C \cap W_i|$ the {\em leader} of $C$.
	\end{lemma}

	\begin{proof}[Proof of Theorem \ref{thm:greedy_min}]
		Let $\mathcal{C}$ denote the clustering given by \Greedy. For every cluster $W_i$ of OPT, merge all
		the clusters of $\mathcal{C}$ that have $i$ as their leaders.
		Let $\mathcal{C}' = (W'_i)$ be this new clustering. 
		By definition of the greedy algorithm, this operation can only increase the cost since every
		pair of clusters have a negative-majority cut at the end of the algorithm:$\cost{\mathcal{C}} \leq \cost{\mathcal{C}'}.$
		We apply Lemma \ref{lemma:greedy_WW'} to $\mathcal{W}=$OPT and $\mathcal{W'}=\mathcal{C}'$, and obtain:
$\cost{\mathcal{C}'} \leq \cost{\text{OPT}} + n\sum_i |W_i' \setminus W_i|$.
		By definition of $\mathcal{C}'$ we have
		$|W_i' \setminus W_i| = 
				\sum_{\substack{C \in \mathcal{C}: \text{leader}(C) = i}}
				\sum_{j \neq i}{|C \cap W_j|},
		$
		hence
		\begin{equation*}
			\label{eqn:greedy_min3}
			\sum_i |W_i' \setminus W_i| = 
				\sum_{C \in \mathcal{C}}
					\sum_{j \neq \text{leader}(C)}{|C\cap W_j|}
				.
		\end{equation*}
		By Lemma \ref{lemma:greedy_inequality},
$			\sum_{j \neq \text{leader}(C)}{|C \cap W_j|} \leq 2 |\Gamma^+(C \cap W_1, \ldots, C \cap
					W_K)|$.		Finally, to bound OPT from below, we  observe that, for any two clusterings $\mathcal{C}$ and $\mathcal{W}$,
it holds that the sum over $C \in \mathcal{C}$ of ${|\Gamma^+(C \cap W_1, \ldots, C \cap W_K)|} $ is less than
				$\cost{\mathcal{W}}$.
		Combining these inequalities yields the theorem.
	\end{proof}

	\begin{thm}
		Let ALG be a randomized algorithm for \MinDisAgree.
		Then there exists an instance on which ALG has cost at least $n - 1 - \cost{\text{OPT}}$ where OPT is the offline
		optimum. If OPT is constant then $\cost{\text{ALG}} = \Omega(n)\cost{\text{OPT}}$.
		\label{thm:greedy_bound}
	\end{thm}
	\begin{proof}
		Consider two cliques $A$ and $B$, each of size $m$, where all the internal
		edges of $A$ and $B$ are positive. Choose a vertex $a$ in $A$, and a set of
		vertices $b_1, \ldots, b_k$ in $B$. Define the edge labels of $ab_i$ as
		positive, for all $1 \leq i \leq k$ and the rest of the edges between
		$A$ and $B$ as negative.
		Define an input sequence starting with $a, b_1, \ldots, b_k$, followed by
		the rest of the vertices in any order. 

	\end{proof}

	\bibliographystyle{plain}
	\bibliography{mathieu}


\newpage
\strut

\end{document}